# TSformer: A Non-autoregressive Spatial-temporal Transformers for 30-day Ocean Eddy-Resolving Forecasting

Guosong Wang[1], Min Hou[2], Mingyue Qin[1], Xinrong Wu[1]*, Zhigang Gao[1], Guofang Chao[1], Xiaoshuang Zhang[1]

[1]National Marine Data and Information Service, Tianjin, China.

[2]Tianjin Binhai New Area Meteorology Administration, Tianjin, China.

Corresponding author: Xinrong Wu (xrw_nmdis@163.com)

**Key Points:**

- A novel non-autoregressive spatial-temporal transformer is developed for rapid 30-day ocean eddy-resolving forecasting
- By extending a temporal dimension, TSformer maintains 3D consistency in physical movements, ensuring long-range coherence.
- TSformer effectively simulates the vertical cooling and mixing effects induced by Super Typhoon Saola

**Abstract**

Ocean forecasting is critical for various applications and is essential for understanding air-sea interactions, which contribute to mitigating the impacts of extreme events. State-of-the-art ocean numerical forecasting systems can offer lead times of up to 10 days with a spatial resolution of 10 kilometers, although they are computationally expensive. While data-driven forecasting models have demonstrated considerable potential and speed, they often primarily focus on spatial variations while neglecting temporal dynamics. This paper presents TSformer, a novel non-autoregressive spatiotemporal transformer designed for medium-range ocean eddy-resolving forecasting, enabling forecasts of up to 30 days in advance. We introduce an innovative hierarchical U-Net encoder-decoder architecture based on 3D Swin Transformer blocks, which extends the scope of local attention computation from spatial to spatiotemporal contexts to reduce accumulation errors. TSformer is trained on 28 years of homogeneous, high-dimensional 3D ocean reanalysis datasets, supplemented by three 2D remote sensing datasets for surface forcing. Based on the near-real-time operational forecast results from 2023, comparative performance assessments against in situ profiles and satellite observation data indicate that, TSformer exhibits forecast performance comparable to leading numerical ocean forecasting models while being orders of magnitude faster. Unlike autoregressive models, TSformer maintains 3D consistency in physical motion, ensuring long-term coherence and stability in extended forecasts. Furthermore, the TSformer model, which incorporates surface auxiliary observational data, effectively simulates the vertical cooling and mixing effects induced by Super Typhoon Saola.

**Plain Language Summary**

The oceans, which cover about 70% of our planet's surface, play a huge role in how our climate works because of their ability to store a lot of heat. They can also be the source of dangerous events like storms, typhoons, and giant waves, which can cause a lot of harm and damage. Being able to predict what the ocean will do quickly and accurately is really important but also very hard. Traditional ocean numerical forecasting systems rely on hundreds to thousands of CPU nodes, requiring 3 to 6 hours of processing time to simulate the 3D ocean for the next 10 days. Recently, data-driven forecasting models based on the latest artificial intelligence technology have surpassed traditional numerical methods in the accuracy of medium and long-term weather forecasting, and the computing speed has been increased by more than 10,000 times. However, due to these intelligent forecasting models focusing more on spatial variation characteristics and neglecting temporal variation characteristics, this may lead to an increase in cumulative errors, thereby affecting the accuracy of long-term forecasts. Here we introducing a new model that uses a special kind of artificial intelligence called a non-autoregressive spatiotemporal Transformer. This model can achieving 30-day ocean eddy resolution forecasting. Compared with in situ profiles and satellite observations, our model does a better job at forecasting the temperature below 20 meters deep than the best operational ocean forecasting systems we have now. It's also good at helping us manage extreme events, such as successfully forecasting the SST cooling effect caused by Super Typhoon Saola.

# 1 Introduction

The ocean, constituting approximately 70.8% of Earth's surface, is the principal recipient of solar radiation. It facilitates the transfer of energy, heat, salt, carbon, and nutrients through seawater movement, resulting in significant marine phenomena such as mesoscale eddies, which greatly influence marine life distribution and connectivity (Suthers et al., 2011). With a specific heat capacity four times that of air, the ocean absorbs 93% of the heat generated by the greenhouse effect, transferring it to the deep ocean (Cheng et al., 2019). The IPCC AR6 report affirms the global average sea surface temperature (SST) increased by 0.88°C between 1850-1900 and 2011-2020, with 0.60°C of this warming having occurred since 1980 (Fox-Kemper et al., 2021). Additionally, the complexity of the marine environment is manifested not only in the long-term effects of climate change but also in the frequency of marine disasters, such as tropical cyclones (including typhoons and hurricanes), internal waves, and marine heatwaves (X. Liu et al., 2023). Ocean forecasting is essential for addressing climate change, predicting extreme events, and providing a scientific foundation for tackling global challenges (Burnet et al., 2014).

Over the past few decades, the accuracy of operational ocean forecasting has continually improving due to advancements in high-performance computing (HPC) and data assimilation methods (Blockley et al., 2014). Since the inaugural global simulations that achieved eddy resolution and visualized global ocean circulation in 1988 (Semtner Jr & Chervin, 1988), contemporary global operational numerical forecasts now span a range of scales, from weather-scale resolutions of one kilometer to seasonal forecasts with resolutions in the tens of kilometers. These forecasts are facilitated by ocean numerical models such as Nucleus for European Modelling of the Ocean (NEMO) (Gurvan et al., 2017), which integrate diverse datasets including in situ profile data, altimeter data, surface temperature data, and sea ice observations. At present, the leading operational Global Ocean Forecasting Systems(GOFSs), such as the Mercator Ocean Physical System (PSY4) and the Real-Time Ocean Forecast System (RTOFS), use physics-driven models in fluid mechanics and thermodynamics with HPC to predict future ocean motion states and phenomena based on current ocean conditions. These systems cover global-to-coastal marine environments and physical and biogeochemical properties, with forecasts typically extending up to 10 days in advance (Blockley et al., 2014).

The augmentation of HPC capabilities has facilitated an escalation in the horizontal resolution of global ocean models, transitioning from 1/10° to 1/32°. This enhancement has led to substantial improvements in the simulation of critical oceanographic phenomena, including surface eddy kinetic energy, the principal pathways of the Kuroshio and Gulf Stream currents, and global tidal dynamics(Guo et al., 2024). Compared to their lower-resolution counterparts, high-resolution models within the Community Earth System Model are now capable of directly simulating small to medium-scale atmospheric and oceanic extreme phenomena, such as tropical cyclones, ocean eddies, and frontal systems (S. Zhang et al., 2020). However, the computational and operational demands for ocean simulations at appropriate spatial and temporal scales are substantial, necessitating HPC to deliver forecasts and services within practical timeframes. Research indicates that elevating the resolution from 1/10° to 1/32° results in an approximate increase in computational load and memory overhead by a factor of 32 and 10, respectively (B. Xiao et al., 2023), posing significant technical challenges for model development and operational efficiency.

On the other hand, deep learning technology, characterized by its rapid computational prowess, has furnished ocean forecasting with more robust tools and methodologies. Over the past decade, the exponential expansion of spatiotemporal earth observation and reanalysis datasets has catalyzed the emergence of data-driven models that harness deep learning (Li et al., 2021). These models are showing remarkable potential across a range of earth system forecasting tasks, such as nowcasting of extreme precipitation (Ravuri et al., 2021; Y. Zhang et al., 2023), ocean forecasting (Berbić et al., 2017; Wolff et al., 2020), climate predictions (Ham et al., 2019; Weyn et al., 2021), and ocean phenomenon recognition (Ashkezari et al., 2016). Prior research has amalgamated Recurrent Neural Networks (RNN) and Convolutional Neural Networks (CNN) to leverage temporal and spatial inductive biases, effectively capturing spatiotemporal patterns (Shi et al., 2015; Y. Wang et al., 2019). CNN methods, utilizing satellite observations and gridded Array for real-time geostrophic oceanography Project (Argo) data, have been extensively employed in the reconstruction and prediction of long-lead monthly three-dimensional ocean temperature (C. Xiao et al., 2022), as well as in forecasting surface 2D ocean environmental factors such as SST and Sea-Level Anomaly (SLA) (G. Wang et al., 2022). These intelligent identification and forecasting methods exhibit significant advantages in terms of computational efficiency and predictive accuracy over conventional methodologies (Dong et al., 2022).

Recently, the Transformers and their variants have exhibited exceptional performance across a spectrum of computer vision tasks (Dosovitskiy et al., 2021; Vaswani et al., 2017). Renowned for their exceptional parallel computing capabilities and ability to capture long-range dependencies, Transformers have made it feasible to train extremely large parameter models. Notably, the advanced Transformer-based forecasting models, has become shockingly good at forecasting the weather while using way fewer resources than numerical modeling systems(Han et al., 2024). For example, FourCastNet, which employs the Adaptive Fourier Neural Operator architecture, can generate medium-range weather forecasts globally with an accuracy nearing state-of-the-art, while being five orders-of-magnitude faster than physics-based numerical weather prediction (Kurth et al., 2023). GraphCast, trained on reanalysis data, predicts hundreds of weather variables for the next 10 days at a 0.25° global resolution within under 1 minute, enhancing severe event prediction, including tropical cyclone tracking, atmospheric rivers, and extreme temperatures(Lam et al., 2023). Pangu-Weather, incorporating a 3D Earth-specific transformer architecture and hierarchical temporal aggregation strategy, yields superior deterministic forecast results on reanalysis data compared to the European Centre for Medium-Range Weather Forecasts (ECMWF) Integrated Forecast System (IFS), the world's leading numerical weather prediction system, while also achieving significantly faster computational performance (Bi et al., 2023).

In comparison to computer vision tasks, 3D ocean variables, such as 3D temperature and salinity (3D TS), present higher dimensionality and resolution, and encompass more intricate physical processes. The representation of these 3D variables necessitates a significantly larger number of input tokens. The global self-attention mechanism employed in the Transformer architecture is impractical for data with high dimensionality and resolution due to its computational complexity, which scales quadratically with the size of the 3D data. To address this, the Swin Transformer incorporates a window-attention mechanism (Z. Liu et al., 2021) , which segments the input tensor into non-overlapping local windows and computes self-attention independently within each window, thus circumventing the quadratic complexity associated with global self-attention (Gao et al., 2022). The FuXi model, utilizing 48 repeated Swin Transformer V2 blocks(Z. Liu, Hu, et al., 2022), delivers 15-day global forecasts with a temporal resolution of 6 hours and a spatial resolution of 0.25°(Chen et al., 2023). Furthermore, FuXi-S2S demonstrates an enhanced ability to capture forecast uncertainty and accurately predict the Madden-Julian Oscillation (MJO), extending the skillful MJO prediction from 30 to 36 days (Chen et al., 2024). Additionally, the XiHe model, a data-driven global ocean eddy-resolving forecasting model with a 1/12° resolution, employs a hierarchical Swin-transformer-based framework coupled with a land-ocean mask mechanism and ocean-specific blocks to effectively capture both local ocean information and global teleconnections (X. Wang et al., 2024). These advancements in data-driven forecasting models have resulted in valuable tools for identifying precursor signals, providing researchers with insights, and potentially heralding a new paradigm in earth system science research.

However, training a large-scale Swin Transformer model for high-resolution ocean forecast reveals several issues, including training instability. Modeling the spatiotemporal dynamics of 3D ocean variables presents a significant challenge for deep learning architectures. Current data-driven forecasting models predominantly emphasize the 3D spatial aspects of ocean data through the deployment of 3D neural networks. These models often utilize autoregressive techniques for temporal forecasting to mitigate computational demands, thereby neglecting the system's temporal evolution. Given the chaotic nature of ocean systems, the variability is acutely responsive to both initial spatial states and temporal fluctuations. The efficacy of integrating spatial-temporal attention mechanisms into RNN and Transformer models for these complex systems remains an open question. Beyond minimal adaptation from Swin Transformer, recent studies have incorporated additional inductive biases into the design of space-time Transformers, including trajectory(Patrick et al., 2021), Multiscale Vision Transformers (Fan et al., 2021), and Multiview Transformer(Yan et al., 2022) approaches. However, no prior research has explicitly focused on the development of space-time Transformers for the specific purpose of 3D TS forecasting.

In this study, we introduce TSformer, a novel non-autoregressive spatial-temporal Transformer specifically crafted for medium-range ocean eddy-resolving forecasting of 3D TS, with a forecasting time scale of up to 30 days. The TSformer model is meticulously engineered to efficiently extract complex 3D features and infer relationships from consistent, homogeneous, high-dimensional ocean datasets. Notably, the TSformer model development leverages a 28-year daily ocean physical reanalysis dataset with a spatial resolution of 0.08°. Employing the pre-trained model parameters, we integrated near real-time satellite remote sensing data as surface forcing and utilized the 3D TS nowcast fields as initialization inputs to assess the forecast outcomes for the year 2023. The evaluation, based on in situ and satellite observations, indicates that the TSformer matches with the PSY4 numerical forecast results.

Furthermore, as exemplified by Super Typhoon Saola (2309), the TSformer surpass those of other deep-learning models in accuracy, especially concerning SST cooling response, due to the integration of auxiliary observational data.

**2 Data**

2.1 3D eddy-resolving ocean physical reanalysis Datas

In this study, we employ a global eddy-resolving physical ocean and sea ice reanalysis data (GLORYS12V1) for training and validation in deep learning. The GLORYS12V1 product is the Copernicus Marine Environment Monitoring Service (CMEMS) global ocean eddy-resolving (1/12° horizontal resolution, approximatively 8 km, 50 vertical levels, daily mean) reanalysis covering the altimetry (https://resources.marine.copernicus.eu/product-detail/ GLOBAL_MULTIYEAR_PHY_001_030/),which provides a high-quality and consistent global ocean reanalysis product(Jean-Michel et al., 2021). It utilizes the NEMO ocean numerical model, which is driven at surface by European Centre for Medium Range Weather forecast ERA-Interim then ERA5 reanalysis for recent years. Observations are assimilated by means of a reduced-order Kalman filter. Along track altimeter data, Satellite SST, Sea Ice Concentration and In-situ Temperature and Salinity vertical Profiles are jointly assimilated. Moreover, a 3D-VAR scheme provides a correction for the slowly-evolving large-scale biases in temperature and salinity.

2.2 2D Remote sensing Auxiliary Dataset for surface forcing

The wealth of 2D Ocean satellite remote sensing observations is crucial for advancements in marine science and oceanographic numerical forecasting. This study utilizes three long-term delay-time ocean satellite surface observations as auxiliary data to extract meaningful features from these observations. The multisatellite altimeter SLA dataset is distributed by Archiving, Validation and Interpretation of Satellite Oceanographic (AVISO), which provides a consistent and homogeneous catalog of products. The gridded wind speed (SPD) product are from the Cross Calibrated Multi-Platform (CCMP) V3.1 dataset, which provides Gap-free ocean surface wind data of high quality and high temporal and spatial resolution(Mears et al., 2022). CCMP is a combination of ocean surface (10m) wind retrievals from multiple types of satellite microwave sensors and a background field from reanalysis. The SST data are from the Operational Sea Surface Temperature and Sea Ice Analysis (OSTIA) system(Good et al., 2020), which provides daily gap-free maps of foundation SST and ice concentration at 0.05deg.x 0.05deg. horizontal grid resolution, using in-situ and satellite data.

2.3 Evaluation Dataset

The Argo is an international program that aims to rapidly, accurately, and extensively collect temperature and salinity profile data from the upper layers of the global ocean to improve the accuracy of climate forecasts and effectively defend against the threats posed to humanity by increasingly severe global climate disasters, such as hurricanes, tornadoes, ice storms, floods, and droughts (Wong et al., 2020). In this paper, we utilize the 941 Argo profiles from 2023 that have undergone delayed-mode quality control to assess and evaluate the forecasting capabilities of the TSformer (Figure 1). The process of Argo profiles includes routine quality control, such as duplication removal, landing inspection, climatological boundary examination, spike inspection, and stability testing, as well as salinity drift calibration(Chao et al., 2021).

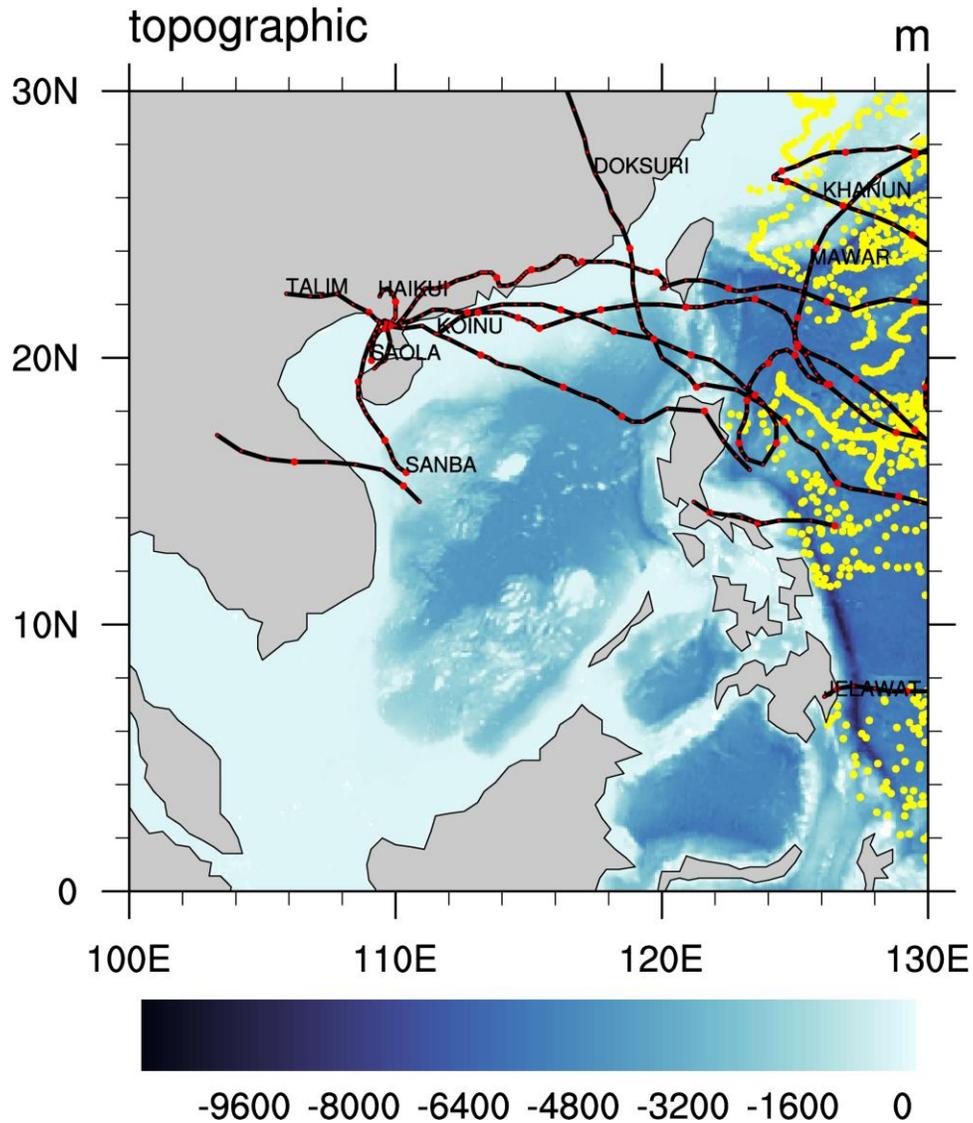

**Figure 1.** The launch location of the 941 Argo floats (yellow points) under delayed mode quality control in the South China Sea (SCS) for the year 2023. The shaded represents the topographic map of the study area, and the black lines indicate the paths of the 20 typhoons that affected the SCS in 2023.

In addition, to further examine the response of SST to typhoons, we have evaluated the SST forecasts based on the Optimally Interpolated (OI) daily SST products using microwave and infrared data (MW_IR) at a 9 km resolution (MW_IR OI SST). The 9 km MW_IR OI SST product combines the through-cloud capabilities of the microwave data with the high spatial resolution and near-coastal capability of the infrared SST data, which can still obtain reliable SST observations even under extreme weather conditions such as heavy rainfall during typhoons(Sun et al., 2018).

2.4 Data pre-processing and Model Domain

This study focuses on forecast 3D TS with a spatial resolution of 1/12° and a vertical resolution of 26 elevation levels, utilizing three 2D Remote Sensing Datasets (SLA, SST, and SPD) as auxiliary data for surface forcing. Consistent with prior research, the dataset is divided into training, validation, and testing sets. The training subset encompasses 10,227 samples from 1993 to 2020, the validation subset includes 730 samples from 2021 to 2022, and out-of-sample near real-time testing is performed with 365 samples from 2023. To enhance the convergence rate of the gradient descent algorithm, the training subset is normalized to the interval [-1, 1] using min-max scaling, with land points assigned a value of zero. The same normalization parameters derived from the training subset are applied to the validation and test subsets.

The spatiotemporal sequence of the 3D TS forecasting problem is designed to predict the most likely future output time steps (Tout) and the H×W×D grid of the 3D TS sequences, based on the current 3D TS sequence and the auxiliary observations at the preceding input time steps (Tin). This mathematical relationship is formalized in Equation (1).

$$3D\ \widehat{TS}_{T+1:T+Tout} = \arg\max_{3D\ TS_{T-Tin:T}} p(3D\ TS_{T-Tin:T} \mid AUX_{T-Tin:T})$$

$$AUX_{T-Tin:T} = \{SLA_{T-Tin:T}, SSTA_{T-Tin:T}, SPDA_{T-Tin:T}\}$$

(1)

We define the sequences of the 3D TS sequence as a series of matrices $3D\ TS_{T-Tin:T} = [3D\ TS_{T-Tin}, 3D\ TS_{T-Tin+1}, ..., 3D\ TS_T]$, where the auxiliary observations are represented as a series of matrices $AUX_{T-Tin:T}$, each contains a set of SLA, SST, and SPD data. The input and output datasets for the 3D TS model are configured with dimensions of [Tin×H×W×D] and [Tout×H×W×D], respectively. Auxiliary data are formatted to the size of [Tin×H×W×Daux], where Tin and Tout denote the input and output time steps, respectively. Considering the large specific heat capacity of the ocean, which results in a relatively slow response of ocean temperature and salinity to external forcing, both Tin and Tout are set to 10 days to adequately capture the spatiotemporal characteristics of the oceanic system. The variables D and Daux correspond to the vertical levels and sea surface satellite variables, with D assigned to 26 and Daux to 3 in this study. H and W signify the dimensions of the regular latitude/longitude grid. All auxiliary data are uniformly processed to a 1/12° spatial resolution using bilinear interpolation.

Given the significant GPU resources necessary for training high-resolution global ocean forecast models—such as Pangu-Weather, which operates at a 1/4° global spatial resolution and requires approximately 16 days on a cluster of 192 NVIDIA Tesla-V100 GPUs during training—this study concentrates on forecasting within the spatial domain of the South China Sea (SCS), spanning 100°-130°E and 0°-30°N, with spatial dimensions (H and W) set to 360 each. The SCS, located south of mainland China, is the largest semi-enclosed marginal sea in the northwestern Pacific Ocean and is known for its frequent typhoons (refer to Figure 1). Due to its complex topography and the influence of strong seasonal monsoons, improving the forecast accuracy of the marine environment in the SCS presents a longstanding and significant challenge (Tuo et al., 2018; Wang et al., 2014).

## 3 Methods

### 3.1 TSformer model architecture

The TSformer model employs a hierarchical U-net encoder-decoder framework that leverages 3D Swin Transformer blocks. This architecture systematically encodes the input sequence into a hierarchy of representations and facilitates forecasting through a coarse-to-fine approach. The TSformer model comprises four primary components, as depicted in Figure 2. Firstly, the extensive input dataset is subjected to dimensionality reduction via joint spatiotemporal 3D Patch Partitioning. Following this, a U-Net structure is employed for downsampling to extract features across multiple scales. The 3D tokens are then passed through a U-shaped encoder-decoder architecture, which is founded on 3D Swin Transformer blocks and incorporates skip connections. The decoder utilizes the multi-scale memory output from the encoder, along with auxiliary inputs, to perform comprehensive spatiotemporal feature learning. Finally, the 3D Patch Merging operation reintegrates the processed sub-regions to reconstitute the original output configuration. Subsequent sections provide detailed information of each component within the TSformer model.

The 3D Patch Partitioning, treats each 3D patch of size 2×3×3 as a token, partitioning the high-dimensional input data into manageable sub-regions to enhance processing efficiency. This partitioning results in the extraction of [$T_{in}/2 \times H/3 \times W/3$] 3D tokens, with each token encompassing a 522-dimensional feature ($2 \times 3 \times 3 \times D + 2 \times 3 \times 3 \times D_{aux}$). Once the 3D tokens are derived from both the 3D TS data and the 2D auxiliary inputs, a linear embedding layer is applied to project the token features into an arbitrary dimension, denoted by C, which represents the base channel width and is set to 256 in this context.

The Encoder, which employs a hierarchical U-Net architecture, is engineered to capture contextual information and extract features from the input 3D tokens across three stages. Except for the initial stage, each stage begins by downsampling the input feature map to reduce resolution, thereby increasing the receptive field to encompass global information. During the second stage, the patch merging layer achieves a 3x spatial downsampling (1/4°×1/4°), and in the third stage, it performs a 4x spatial downsampling (1°×1°). The Encoder is composed of three 3D Swin Transformer blocks and two downsampling layers, which are designed to progressively reduce the spatial resolution of the original 3D ocean data (1/12°×1/12°) while expanding the number of feature maps. This design aids the model in capturing higher-level features. The 3D Swin Transformer block is instrumental in computing correlations among various spatiotemporal elements, enabling the detection of long-term trends, periodicity, and 3D spatial features within the data. Through these three stages, high-level features for small-scale (1/12°×1/12°), medium-scale (1/4°×1/4°), and large-scale (1°×1°) ocean data are sequentially extracted. This methodology facilitates the efficient extraction of multi-scale spatiotemporal features from the 3D ocean data, leading to a comprehensive understanding of the spatiotemporal characteristics within the dataset.

The Decoder, adhering to the U-Net architectural paradigm, is responsible for the incremental upscaling of feature spatiotemporal dimensions. It plays a critical role in merging features from the downsampling path using skip connections. Consisting of three 3D Swin

Transformer blocks and two upsampling layers, the decoder receives multi-scale feature outputs from the corresponding encoder block and the preceding decoder stage. Each upsampling layer skillfully integrates these inputs. The strategic integration of skip connections alongside 3D Swin Transformer blocks within the decoder architecture is essential for reconstructing the original spatial and temporal attributes of the 3D ocean dataset and its auxiliary inputs.

The 3D Patch Merging, occurring in the final stage of the Decoder, is crucial for transforming feature maps into the ultimate 3D forecasting field. It serves as the inverse of the 3D Patch Partition process, meticulously reassembling the processed sub-regions to reconstruct the 3D output with precision.

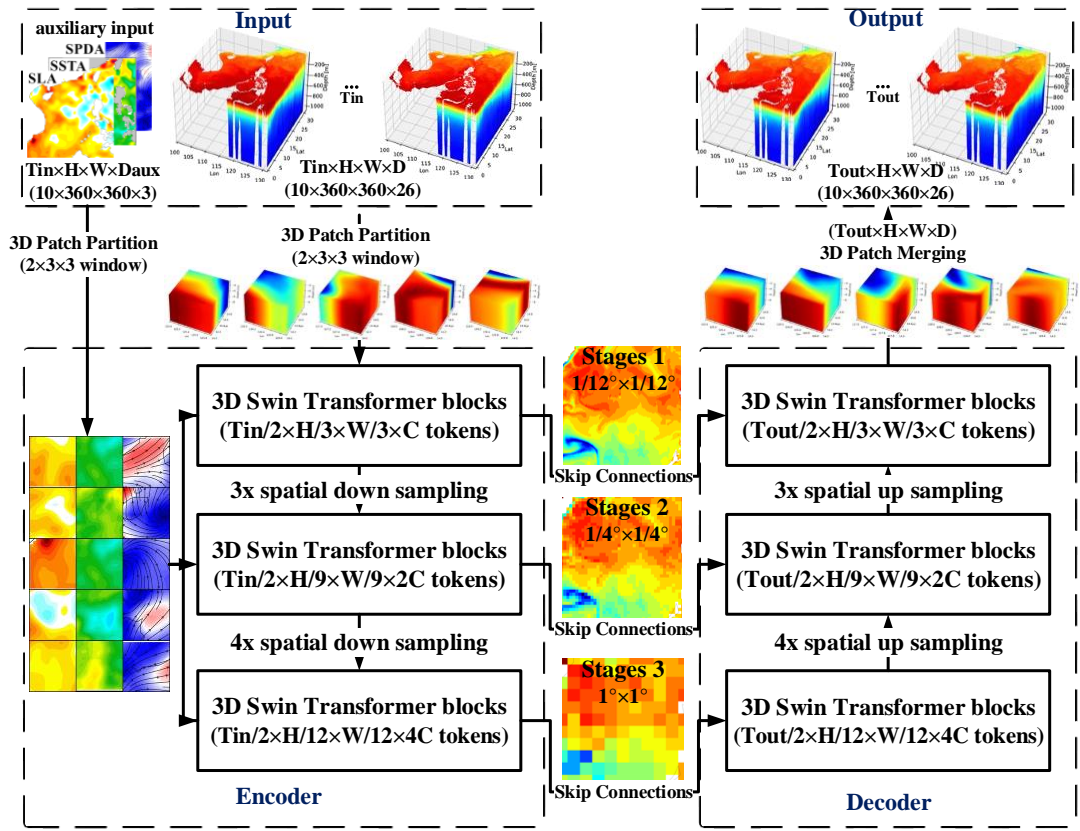

**Figure 2.** An overview of the proposed TSformer model architecture. The TSformer model consists of four main components: 3D Patch Partitioning, Encoder, ecoder and 3D Patch Merging. The input ocean state sequences are defined on a 1/12° latitude-longitude grid, incorporating 10 continuous time series inputs. These inputs include three surface variables as auxiliary data on the left and 3D TS across 26 vertical levels on the right, represented as [Tin×H×W×Daux + Tin×H×W×D]. The 3D Patch Partitioning process reduces the dimensionality of the large input dataset through joint spatiotemporal partitioning, yielding [Tin/2×H/3×W/3×C] 3D tokens. The Encoder component of the TSformer maps these local regions of 3D tokens into three stages of 3D Swin Transformer blocks, which are designed to capture contextual temporal correlations and extract spatial features at three different resolutions: 1/12°×1/12°, 1/4°×1/4°, and 1°×1°. The Decoder component then upscales the processed multi-scale features back onto the grid representation, adhering to the U-Net architectural paradigm and utilizing skip connections to merge features from the downsampling path. Finally, the 3D Patch Merging operation in the final stage meticulously reassembles the processed sub-regions to reconstruct the 3D output with dimensions [Tout×H×W×D].

3.2 3D Swin Transformer block

We propose the 3D Swin Transformer block(see Figure 3), which is designed to address various data correlations, including temporal and spatial correlations, by employing multiple structure-aware space-time attention layers. This methodology extends the scope of local attention from purely spatial to encompass spatiotemporal computations. Consequently, the TSformer is capable of capturing the intricate temporal dynamics among different variables and extracting fine-grained spatial features across various vertical layers. This dual capability significantly bolsters the model's forecasting accuracy.

The computation formula of the 3D Swin Transformer block is shown in Equation (2).

$$\begin{aligned}
\hat{A}^l &= 3D\text{-}W\text{-}CA(LN(A^{l-1})) + A^{l-1}, \\
A^l &= FFN(LN(\hat{A}^l)) + \hat{A}^l, \\
\hat{X}^l &= 3D\text{-}W\text{-}MSA(LN(X^{l-1})) + X^{l-1}, \\
X^l &= FFN(LN(\hat{X}^l)) + \hat{X}^l, \\
\hat{X}^{l+1} &= 3D\text{-}SW\text{-}MSA(LN(concat(X^l, A^l))) + concat(X^l, A^l), \\
X^{l+1} &= FFN(LN(\hat{X}^{l+1})) + \hat{X}^{l+1},
\end{aligned} \quad (2)$$

where $\hat{A}^l$ and $\hat{X}^l$ denote the output features of the 3D Window-based Cross-Attention (3D-W-CA) module and the 3D Window-based multi-head self-attention (3D-W-MSA) module for block $l$, respectively. $A^l$ and $X^l$ denote the output features of the Feed-forward network (FFN) module for block $l$, respectively. $\hat{X}^{l+1}$ and $X^{l+1}$ denote the output features of the the 3D Shifted Window-based multi-head self-attention (3D-SW-MSA) module and FFN module for block $l+1$, respectively.

In the 3D Swin Transformer block, the 3D TS and auxiliary data are divided into spatiotemporal cuboid patches to engage the 3D-W-CA and 3D-W-MSA modules, respectively (Z. Liu, Ning, et al., 2022). The 3D-W-CA module computes the correlation and weighted sum between distinct sequences of the 3D TS and auxiliary input tokens, embodying an advanced form of multi-head attention. Conversely,

. In contrast, the 3D-W-MSA module analyzes the association between each element within the same input 3D TS sequence, revealing intrinsic features of the 3D TS to facilitate long-term forecasting capabilities. Then, the attended weights from the FFN module are concatenated to reconstruct the output feature maps. This process enables the 3D Swin Transformer block to capture the complex spatiotemporal patterns within the 3D TS and auxiliary data through 3D-W-CA and 3D-W-MSA module. Finally, the 3D-SW-MSA module incorporates a shift window to introduce cross-window connections between adjacent non-overlapping 3D cuboid windows from the previous layer.

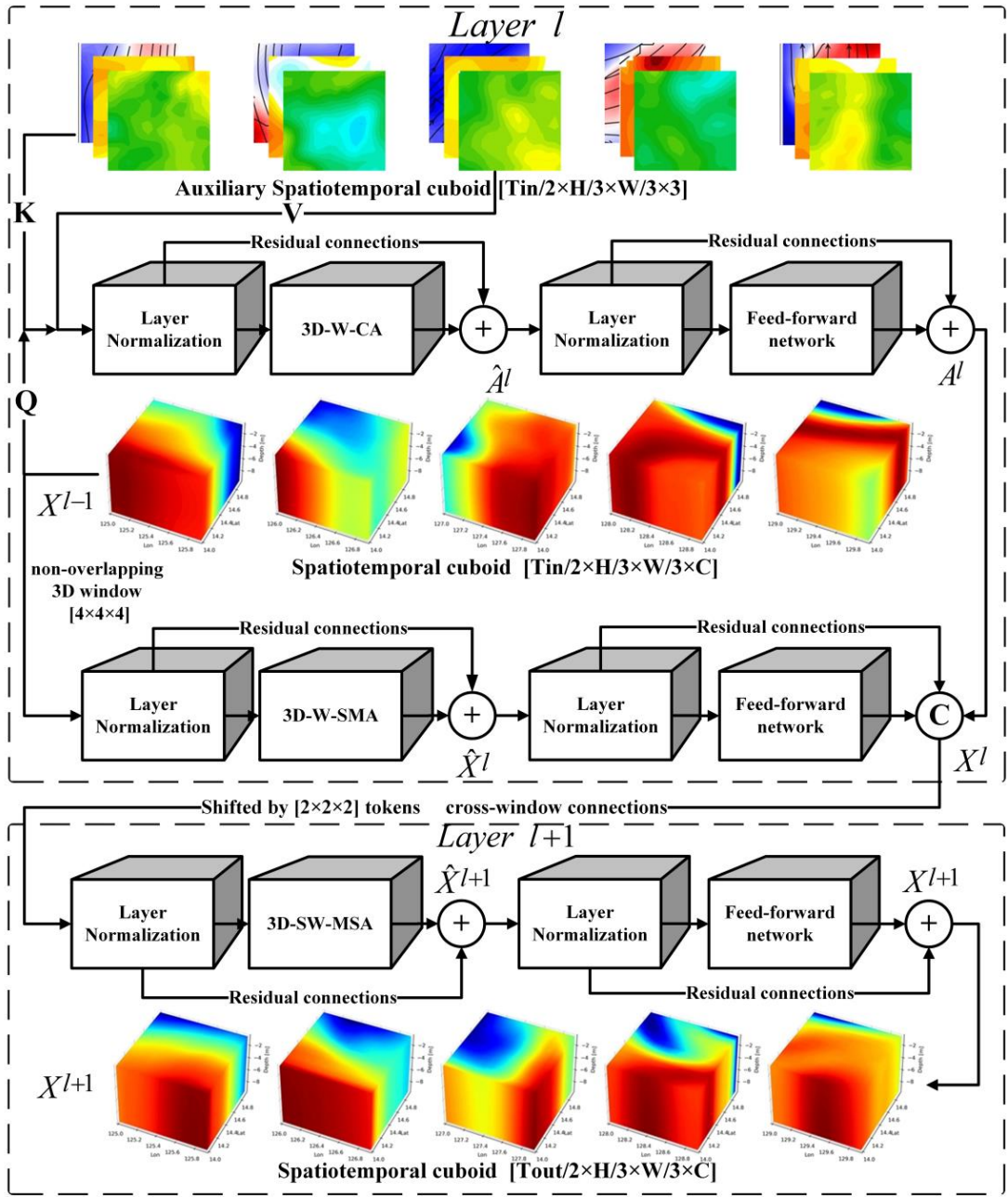

**Figure 3.** An illustration of two successive 3D Swin Transformer block. The 3D Swin Transformer block used in the paper is composed of three main units: The first unit consists of a 3D-W-CA module followed by a FFN with GELU non-linearity activation function. In the context of CA, Q is derived from one sequence (e.g., the ocean model's internal state), while K and V are extracted from the other sequence (e.g., SST, SLA, and SPD). This allows the model to incorporate information from the auxiliary data into its forecast of the 3D TS. The second and third unit is symmetric to the first unit, with the only difference being the replacement of the 3D-W-CA module with a 3D-W-MSA module and a 3D-SW-MSA module, respectively. The other components remain the same. Layer Normalization (LN) is applied before each 3D-W-MSA or 3D-SW-MSA module and FFN, this layer normalizes the input 3D tokens to have zero mean and unit variance, improving the stability and convergence of the network. Residual connections are applied after each module, which means the input is added to the output of the module. This helps in the flow of gradients during training and improves the overall performance of the network.

### 3.3 Non-autoregressive methods

Purely data-driven machine learning models, in contrast to physics-based numerical models, often lack the incorporation of physical constraints, which can lead to significant error propagation and unrealistic predictions over long lead times (Lam et al., 2023). When making iterative forecasts, error accumulation is unavoidable as lead times increase. To address this challenge, the TSformer model is designed as a non-autoregressive model, allowing it to directly forecast multiple time steps ahead without relying on autoregressive methods that utilize its own single-step forecast as input. This capability enables the generation of arbitrarily long lead times of ocean states. Typically, the TSformer utilizes a sequence of 3D TS data and auxiliary observations from the previous 10days steps to forecast 3D TS data for the next 10days steps. Notably, generating 30-day forecasts using a single TSformer model necessitates only three iterative runs. This approach is analogous to expanding the assimilation time window in four-dimensional variational data assimilation (4D-Var) methods (Courtier et al., 1994; Y. Xiao et al., 2023), which aim to optimize initial state estimation by minimizing discrepancies between model forecasts and observations (labels) over a specified assimilation time window(Dee et al., 2011; Lorenc & Rawlins, 2005).

### 3.4 Train and Hyperparameters

The TSformer model, developed utilizing the PyTorch Lightning framework, comprises approximately 222 million parameters, with hyperparameters detailed in Table 1. The pretraining phase of the TSformer is anticipated to last around 10 days, utilizing a cluster of 8 Nvidia A800 GPUs. We employ the AdamW optimizer(Llugsi et al., 2021), setting $\beta_1$ to 0.9 and $\beta_2$ to 0.999, and incorporate a negative slope of 0.1 for the LeakyReLU activation function. The training process spans 200 epochs across all datasets, with early stopping initiated based on the validation score, permitting a patience of 10 epochs. A 20% linear warm-up phase precedes the Cosine learning rate scheduler, which gradually decreases the learning rate to zero after the warm-up period. To mitigate memory consumption, Fully-Sharded Data Parallel(Y. Zhao et al., 2023) is implemented during training. Considering the high-resolution and high-dimensional nature of the dataset, the TSformer may require a larger parameter set and extended training durations. Therefore, practical deployment might require customized adjustments and optimizations to align with specific task demands and resource constraints.

**Table 1.** Hyperparameters for training the TSformer model

| Hyperparameters | Value |
| --- | --- |
| Inputs Size | 10×360×360×26<br>10×360×360×3 |
| Output Size | 10×360×360×26 |
| Loss Function | RMSE |
| Optimizer | AdamW |
| Learning Rate | 0.001 |
| $\beta_1$ | 0.9 |
| $\beta_2$ | 0.999 |
| Batch Size | 16 |
| Weight decay | 0.00001 |
| Learning rate decay | Cosine |
| Max Training Epochs | 200 |
| Warm up percentage | 10% |
| Early stop | True |
| Early stop patience | 10 |
| Parameters | 222 million |

**4 Operational Forecast Results**

To evaluate the operational forecasting performance of the TSformer model, we integrated near real-time satellite remote sensing data as surface forcing and utilized the 3D TS nowcast fields from the Operational Mercator global ocean analysis and forecast system , which is part of the real-time global forecasting CMEMS system(Lellouche et al., 2018). This system closely mirrors the GLORYS12 reanalysis dataset (Jean-Michel et al., 2021). These fields were used as initialization inputs for our model throughout 2023.

The TSformer operates daily, providing forecasts from January 1, 2023, to December 31, 2023, offering forecast leading to 30 days. The physical 30-day forecast products from the TSformer, hereafter referred to as the 3D TS forecast results, include daily mean 3D TS fields on standard 1/12° grid (0.0833° latitude x 0.0833° longitude) in the SCS, with 26 geopotential levels ranging from 0 to 1000 m, consistent with the resolution of the GLORYS12 reanalysis data.

The accuracy of the 3D TS forecast results was evaluated using the 2023 evaluation dataset. The assessment involved out-of-sample 3D TS reanalysis datasets, quality-controlled Argo observations at delayed times (reprocessed when available), and independent WM_IR SST data. It is essential to underline that the uncertainty in analyzing SST and observations is higher in near real-time forecasts compared to hindcast runs.

4.1 Metrics

We objectively assess the accuracy and skill of the 3D TS forecasts from two distinct perspectives. Accuracy is determined by the discrepancy between the forecast and observations or analyses, while skill is evaluated by comparing the forecast performance to a reference method, such as persistence or an alternative forecast system. The persistence model posits that the initial forecast state remains unchanged throughout the entire lead time (Levine & Wilks, 2000), and it represents a cost-effective forecasting approach (Shriver et al., 2007). The alternative forecast systems used in this paper includes the PSY4 numerical forecast system, the TSformer model without auxiliary observational data (TSformer-w/o-aux), and the TSformer model integrated with autoregressive methods (TSformer-AR).

The evaluation was conducted using three principal performance indicators: Bias, Root Mean Square Error (RMSE), and Anomaly Correlation Coefficient (ACC). Bias indicates the presence of systematic errors, a perfect score of Bias=0 does not preclude the possibility of large errors with opposite signs that may cancel each other out, thus the concurrent use of

RMSE is essential for a comprehensive assessment. RMSE, a widely accepted measure of accuracy, shows that higher values correspond to poorer forecasting proficiency. On the other hand, ACC is regarded as a skill metric relative to climatology, with higher values indicating superior forecast skill. An ACC value of 0.5 suggests that the forecast errors are comparable to those of a forecast based on climatological averages alone.

$$Bias = \overline{f - o} \qquad (3)$$

$$RMSE = \sqrt{\overline{(f - o)^2}} \qquad (4)$$

$$ACC = \overline{(f - c)(o - c)} \sqrt{\overline{(f - c)^2 (o - c)^2}}^{-1} \qquad (5)$$

In this context, f denotes the forecast value, while o represents the observed or analyzed value. The climate value, indicated by c, is defined as the long-term average conditions of the ocean over a specified reference period. For the purposes of this paper, the reference period extends from 1993 to 2010. The over-bar symbol (-) signifies an average computed over an extensive sample, encompassing both temporal and spatial dimensions.

4.2 3D TS forecast results Evaluation with GLORYS12V1

Given that current remote sensing satellites are limited to monitoring only the ocean surface conditions, and with underwater temperature and salinity observations heavily dependent on the sparse data from Argo profiling floats (as referenced in Figure 1), we initially employ the GLORYS12v1 reanalysis data as a benchmark for qualitatively evaluating the accuracy of the TSformer model. Through a comparison of the 3D TS forecasts with the GLORYS12v1 reanalysis dataset (Figure 4), it is observed that the TSformer model accurately captures the characteristics of 3D TS variations at various depths. The spatial distribution of the 3D TS forecast across the mixed layer, thermocline, and deep layer consistently matches the GLORYS12v1 data. Upon analyzing the vertical distribution of biases, temperature biases are primarily concentrated in the thermocline region, while salinity biases are predominantly found in the surface layer, with minimal horizontal biases overall. Specifically, the 3D temperature bias is most pronounced in the Luzon Strait and the eastern sea of the Philippines, regions that significantly influenced by water mass distribution and seasonal changes due to the Kuroshio (Qu, 2000). Conversely, the 3D salinity bias is more significant in the northern coastal and southern regions of the SCS, associated with external freshwater inputs and the marine processes driven by the entire SCS monsoon system (Yi et al., 2020).

Furthermore, it is notable that the TSformer model exhibits intriguing emergent capabilities when trained on a large scale and using non-autoregressive methods. These capabilities enable the TSformer to accurately replicate the physical properties and processes of 3D TS data, which are derived from sequences of physical reanalysis datasets. Impressively, even in the absence of explicit inductive biases specific to 3D currents fields, the TSformer is capable of generating 3D TS representations that incorporate dynamic currents patterns. As the movement of water masses, TS elements consistently flow through 3D space as well.

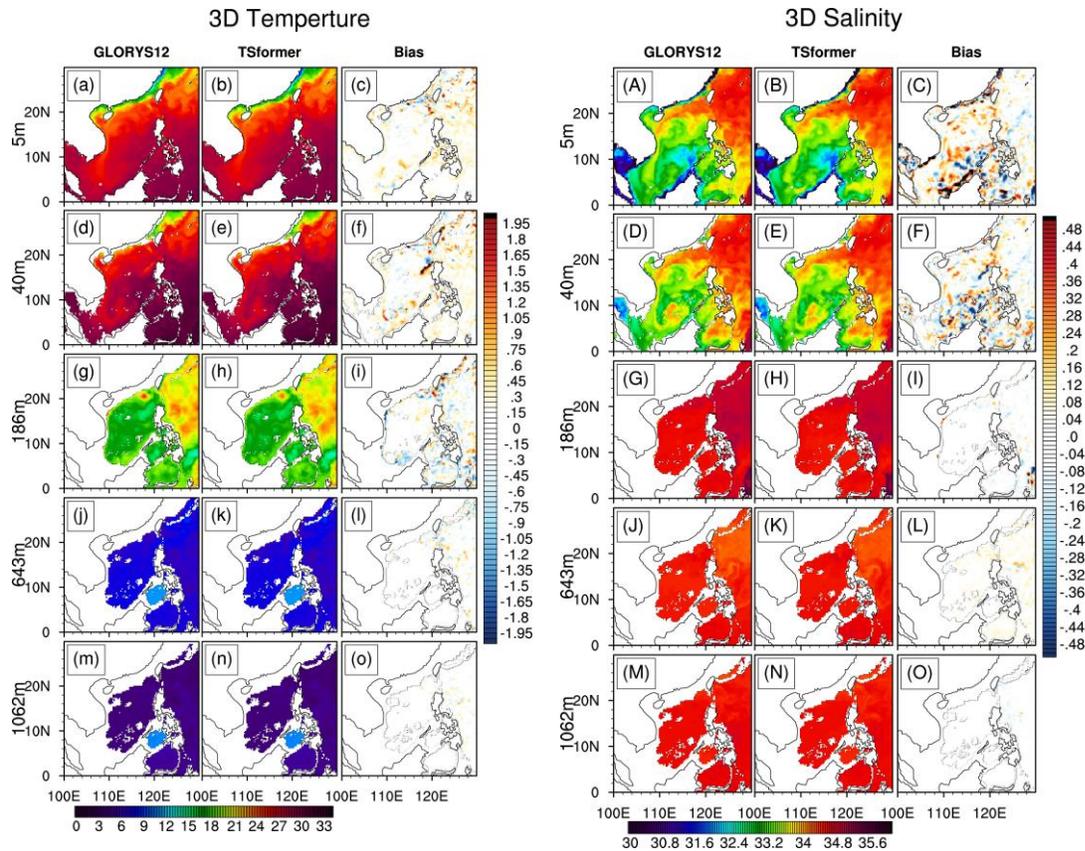

**Figure 4.** An Visualization example of the spatial distribution of forecast performance on January 5, 2023. The forecasts generated by the TSformer model for the fifth day (starting from January 1, 2023) are depicted in the second and fifth columns, whereas the GLORYS12V1 reanalysis data is shown in the first and fourth columns. Additionally, the figure includes a spatial map illustrating the bias between the TSformer and GLORYS12V1 for 3D Temperature (third column) and 3D Salinity (sixth column). The rows of the figure correspond to different depths: the first row represents 5 meters, the second row 40 meters, the third row 186 meters, the fourth row 643 meters, and the fifth row 1062 meters.

Figure 5 illustrates the 10-day average RMSE and Bias over time and depth of the 3D TS forecast. The TSformer model displays an average 3D Temperature Bias of 0.02°C, maintaining a neutral Bias of 0°C at depths below 250 m. Relative to GLORYS12v1, the TSformer exhibits a warm Bias around 30m, reaching a peak of 0.11°C (as depicted in Figure 5A), originating from the initial conditions. At the surface, the average 3D Temperature RMSE is 0.45°C, with the highest RMSE observed at 80m (0.60°C), which subsequently decreases to 0.40°C at depths below 250 m (Figure 5A). For salinity, the maximum RMSE is 0.26 PSU at the surface, with an average RMSE of 0.16 PSU above 100m and 0.04 PSU below 100m (refer to Figure 5B). The average salinity Bias is 0.003 PSU, displaying notable variations with depth.

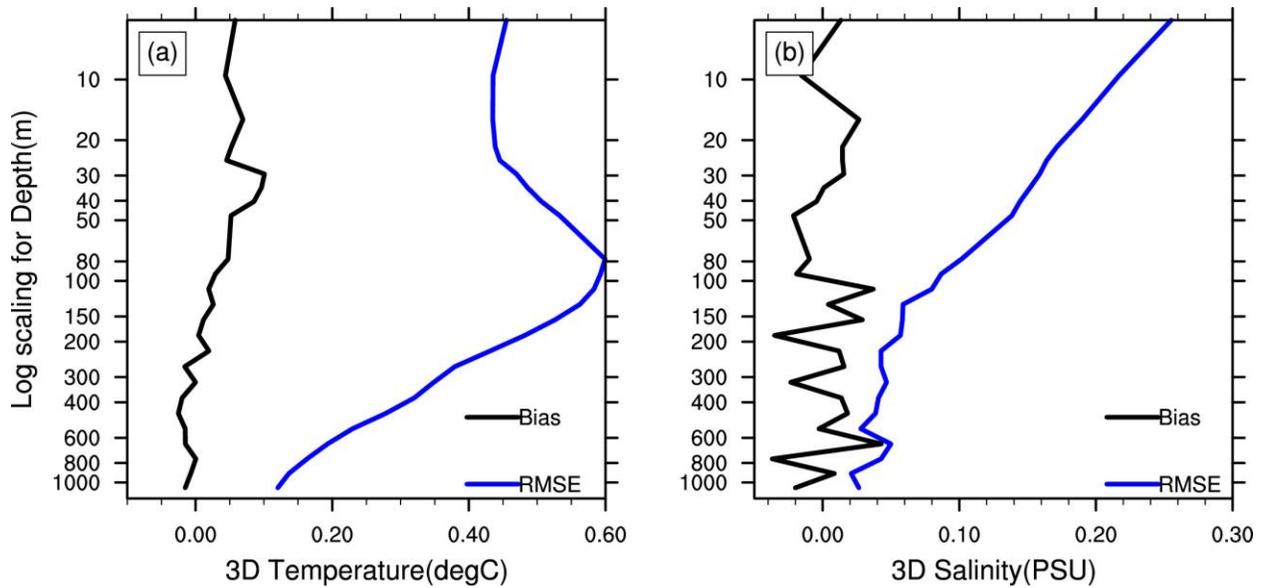

**Figure 5.** The 10-day average RMSE (blue line) and Bias(black line) profiles for the 3D temperature(a) and 3D salinity(b), respectively. Both sets of data derived from the 2023 operational forecast results.

Figure 6 shows the time series comparison of the 10-day average RMSE and Bias between the TSformer model and the GLORYS12v1 reanalysis data, highlighting the good stability of the TSformer model in comparison. Notably, from July to November 2023, there is a certain increase in the RMSE for both temperature and salinity, coinciding with the active typhoon period in the SCS. This phenomenon will be further discussed in Section 4.4.

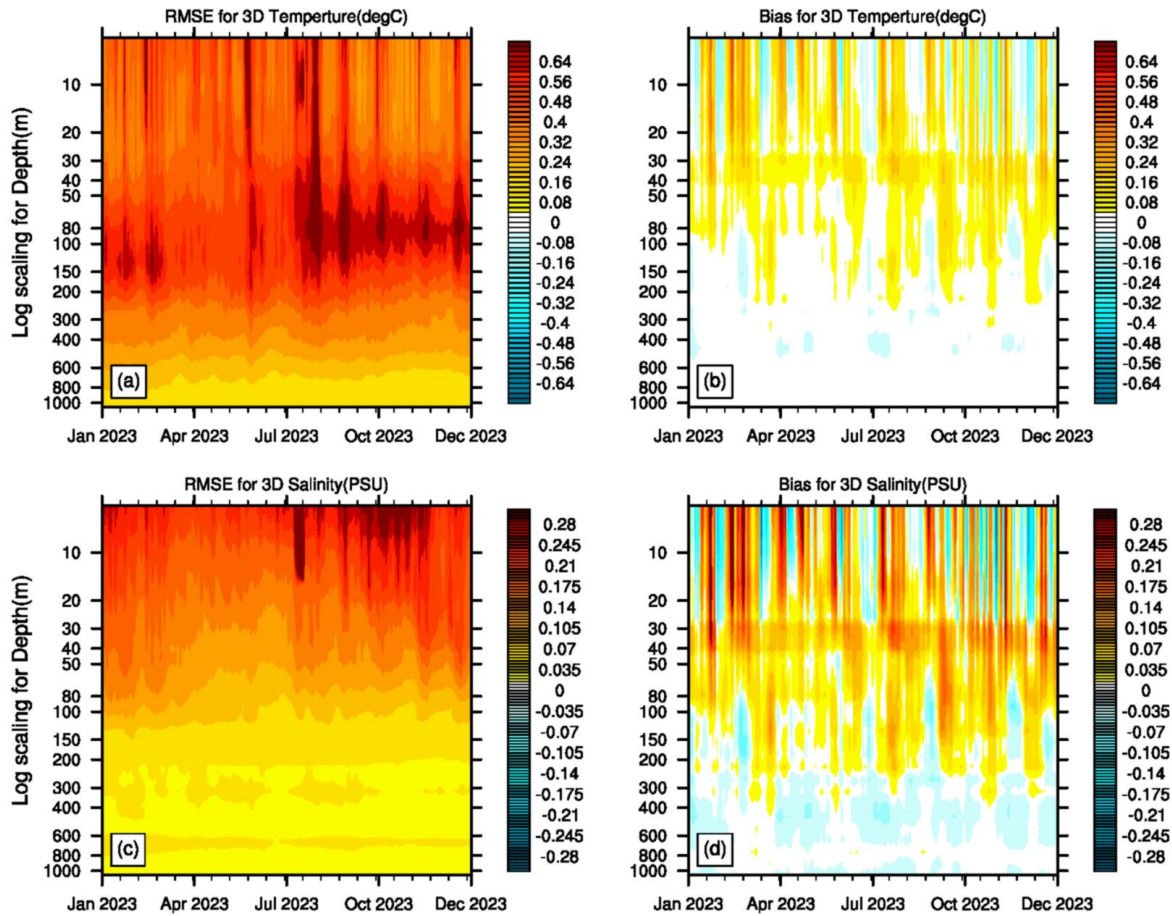

**Figure 6.** The time series comparison of the 10-day average RMSE (first column) and Bias (second column) for the TSformer model and the GLORYS12v1 reanalysis is presented. The comparison is made for the 3D temperature (first row) and 3D salinity (second row), respectively. Both sets of data are derived from the 2023 operational forecast results.

Furthermore, we objectively evaluated the skill of the TSformer model with two reference methods: persistence and the TSformer-AR model, which processes discrete single-time-step tokens as inputs and targets. Figure 7 illustrates the detailed quantitative comparison among these three models based on the 2023 operational forecast results against the GLORYS12V1 reanalysis data over the SCS. At lead times of 2 days, the RMSE and ACC of TSformer-AR are essentially consistent with the TSformer. However, due to its autoregressive methods, the skill of the TSformer-AR declines precipitously beyond 5 cycles, resulting in RMSE values that exceed those of persistence forecasting. Conversely, the TSformer exhibits a gradual and steady increase in RMSE (or decrease in ACC), maintaining an ACC above 0.5 for both temperature and salinity by the 30th day. This performance surpasses all baseline models, and demonstrating superior forecasting capabilities. The enhanced performance of the TSformer can be attributed to its end-to-end training for 3D TS forecasting. By employing an efficient space-time attention block and a U-Net Encoder-Decoder architecture, the TSformer extends the scope of local attention computation from the spatial domain to the spatiotemporal domain. This approach introduces periodic characteristics by expanding the temporal dimension and effectively reduces long-term cumulative errors.

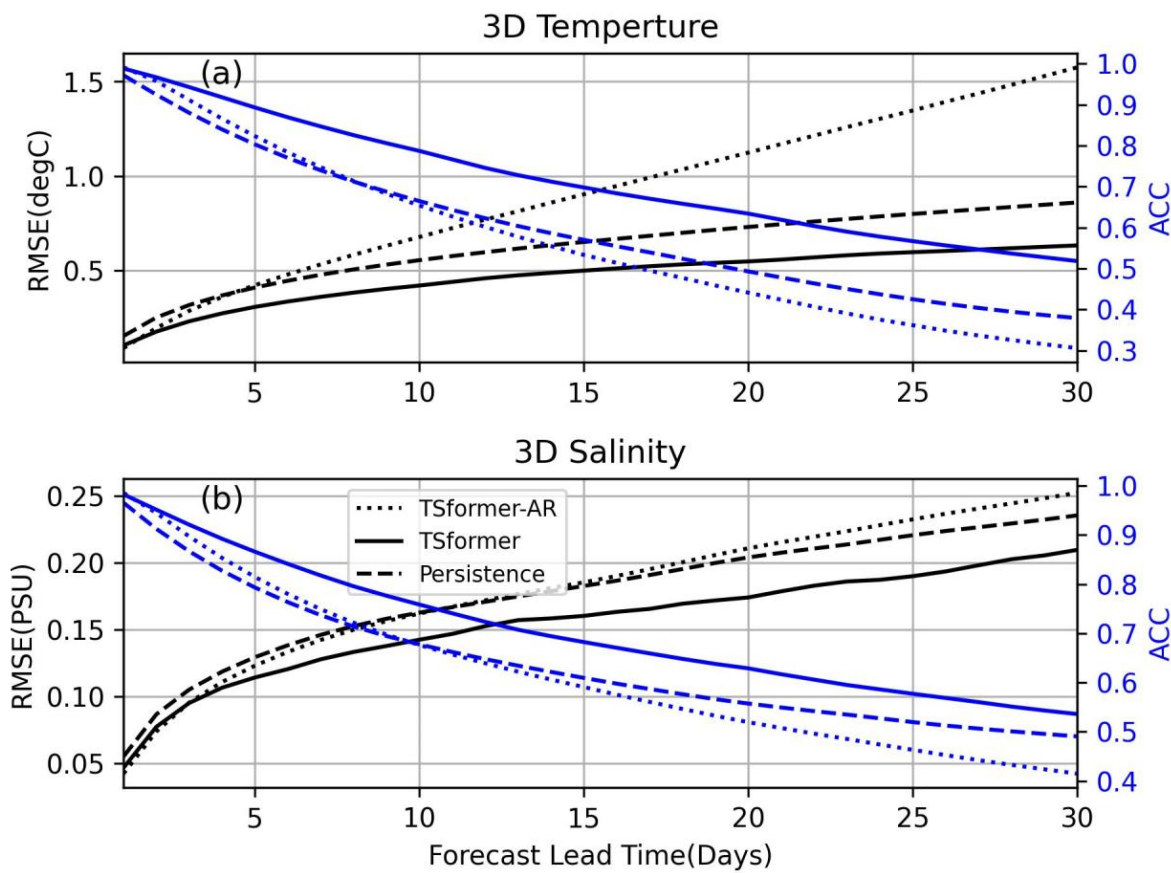

**Figure 7.** The forecast skill comparison of the average RMSE (black line ) and ACC (blue line) based on the 2023 operational forecast results against the GLORYS12V1 reanalysis data over the SCS for both 3D temperature(a) and 3D salinity(b). The x-axis represents the forecast lead time. The dotted, solid, and dashed lines represent the performance of the TSformer-AR (TSformer with AutoRegressive methods), TSformer, and Persistence models (i.e., assuming no change from the initial state), respectively.

4.3 TS vertical profiles Evaluation with Argo

Utilizing the Argo vertical profiles in 2023, the forecasting capability of the TSformer model was evaluated and compared with the state-of-the-art numerical forecasting system PSY4. Figures 8 illustrate the the temperature and salinity RMSE profiles for various forecast lead days at depths exceeding 1000m. It is observed that within the mixed layer, ranging from 0 to 20 meters, the RMSE for TSformer is slightly higher than that of PSY4, with an increase that correlates with the forecast lead time. Specifically, on the first lead day, the difference in 3D temperature RMSE between the two models is negligible (less than 0.05°C), whereas by the 10th lead day, this difference grows to 0.18°C within the mixed layer. This divergence within the mixed layer may arise from the atmospheric field forcings for PSY4, which are derived from the

ECMWF IFS with a 3-hour sampling frequency to capture the diurnal cycle. In contrast, TSformer currently relies on a daily auxiliary data for surface forcing, which has both significantly larger time and spatial intervals compared to PSY4. As a result, TSformer extracts fewer and less comprehensive auxiliary features, leading to increased errors in the surface layer. However, at depths beyond 20 meters, the temperature and salinity forecasts from TSformer significantly surpass those of PSY4. The average maximum RMSE for temperature is observed at a depth of 50m, with TSformer reporting values of 0.89°C and 1.17°C on the first and tenth days, respectively, compared to PSY4 values of 1.23°C and 1.29°C for the corresponding lead times.

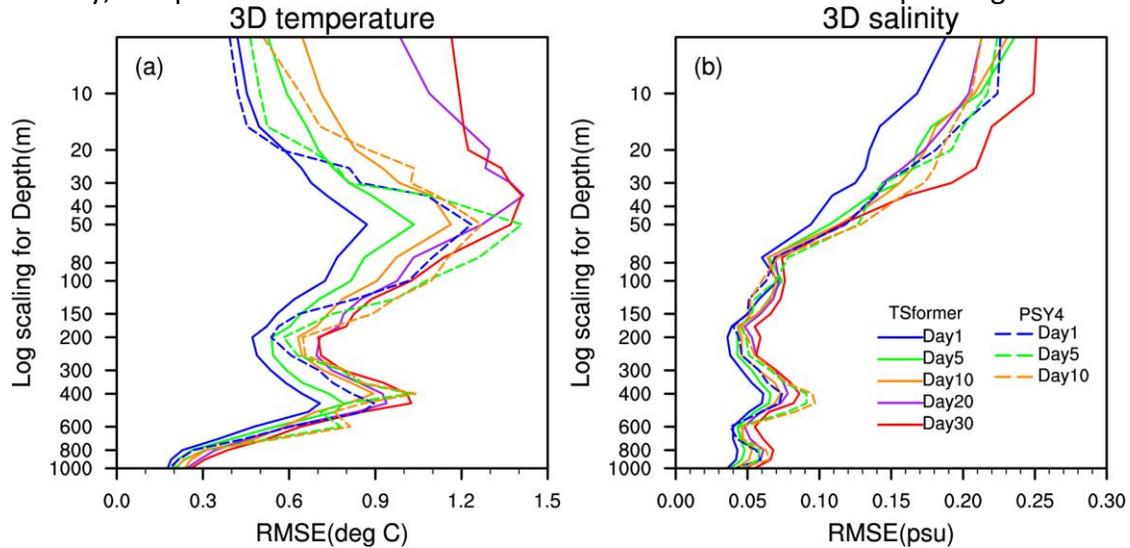

**Figure 8.** The vertical profiles of the average RMSE for 3D temperature (a) and 3D salinity(b), as compared with Argo data for two different models: TSformer (solid line) and PSY4 (dashed line). The lead times for the forecasts are represented by different colors: 1 day (blue), 5 days (green), 10 days (orange), 20 days (purple), and 30 days (red). Both sets of data are sourced from the 2023 operational forecast results.

The time series of the area-weighted RMSE, as depicted in Figure 9, indicates that the TSformer model initiates with RMSE values of 0.59°C for temperature and 0.08 PSU for salinity, both of which are subject to the influence of initial conditions. By the 30th day, these values increase to 0.98°C and 0.12 PSU, respectively. It is noteworthy that, within the initial 10 days of the forecast, the TSformer model matches the performance of the PSY4 model. Unlike PSY4, which relies on HPC for extended numerical simulations, the TSformer is capable of completing a 30-day forecast in approximately 40 seconds utilizing only a CPU, demonstrating a significant advantage in computational efficiency.

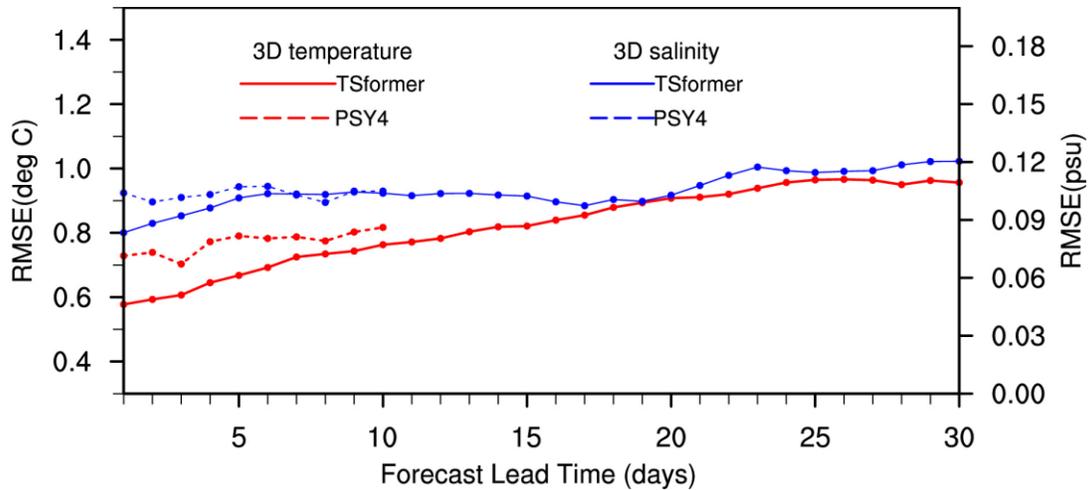

**Figure 9.** The forecast performance of the average RMSE for 3D temperature (red) and 3D salinity(blue), respectively, as compared with Argo data for two different models: TSformer (solid line) and PSY4 (dashed line).The x-axis represents the forecast lead time. The y-axis represents the forecasting RMSE (lower is better). Both sets of data are sourced from the 2023 operational forecast results.

4.4 SST cooling Evaluation with Satellite Observation

The SCS is serves as a critical region for the genesis and landfall of typhoons that originate from the northwest Pacific and the SCS itself(Z. Zhao et al., 2024). In 2023, the SCS encountered 20 typhoons (recall Figure 1), with these events typically occurring between April and December. Additionally, with the annual increase in the heat content of the upper ocean, the intensity of typhoons has shown an upward trend over the past forty years(Guan et al., 2018).

Leveraging the cloud-penetrating capabilities of satellite microwave radiometers, which provide crucial observational data on SST, this study utilizes MW_IR OI SST and compares the TSformer model with the PSY4 model and the TSformer model without auxiliary data (TSformer-w/o-aux) to rigorously evaluate the forecast accuracy and stability under typhoon conditions. The time series analysis of average RMSE and ACC (Figure 10), reveals significant performance differences among the models. Despite utilizing the same 3D TS input data, the TSformer-w/o-aux model, which does not incorporate 2D surface variables as auxiliary input, exhibits lower forecast accuracy for SST compared to the TSformer and PSY4 models, both in average performance and variability. This divergence is especially marked from May to October, with the TSformer-w/o-aux model exhibiting over a 30% increase in RMSE for SST forecasts relative to the TSformer model. Conversely, the TSformer model maintains exceptional stability in SST forecasting, achieving an average ACC of 0.92 and an average RMSE of 0.50°C, which is comparable to the the performance of the PSY4 model. During the typhoon-active months of July to October, the TSformer model achieves a maximum RMSE of 0.81°C for the super typhoon DOKSURI(2305), outperforming the PSY4 model(1.0°C).

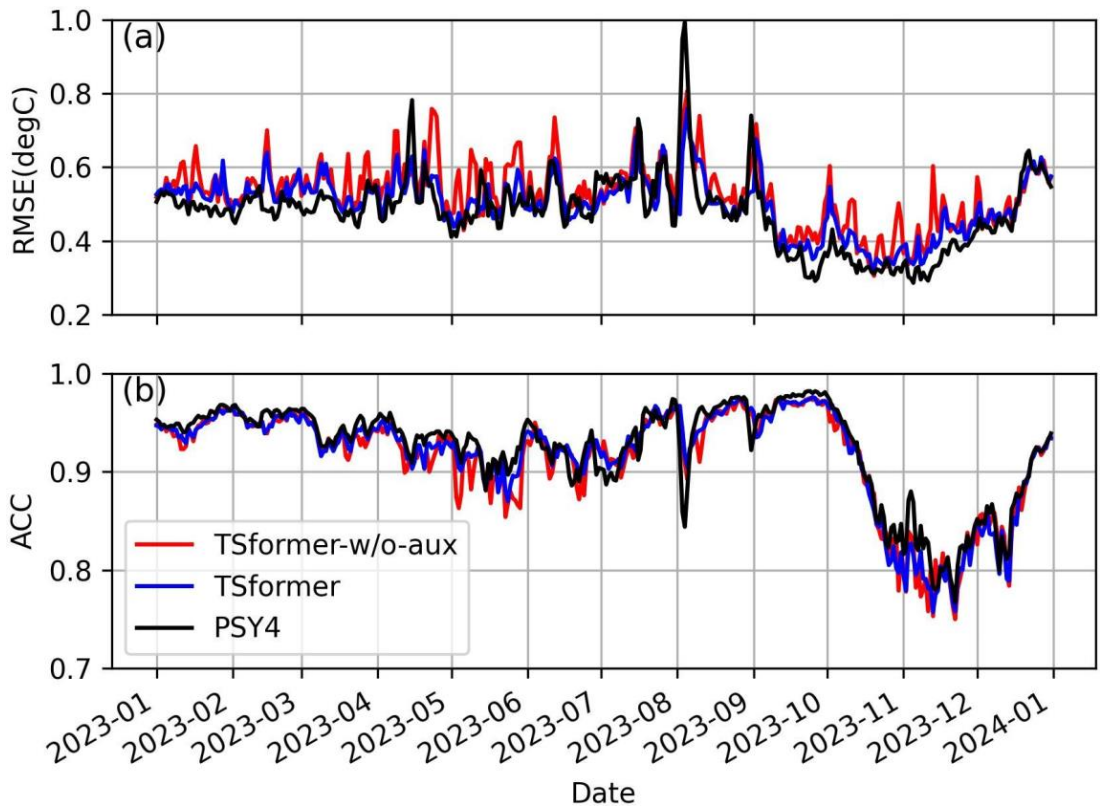

**Figure 10.** The time series comparison of the 3-day average RMSE (a) and ACC (b) of SST forecast is depicted for the TSformer-w/o-aux(red), the TSformer (blue), and the PSY4 (black). These evaluations are benchmarked against the MW_IR OI SST dataset, with both sets of results extracted from the 2023 operational forecast results.

      Tropical cyclones, as intense local disturbances, transfer momentum to the ocean and absorb heat during their movement, leading to significant dynamic and thermal changes in the ocean over a short period(Potter et al., 2017; X. D. Wang et al., 2011). These cyclones induce substantial upper-ocean mixing and upwelling, leading to sea surface cooling, thereby producing a negative feedback effect on the cyclone itself (Jullien et al., 2014). Observations indicated that SST cooling caused by tropical cyclone ranges from 1 to 6 °C(Bender et al., 1993), with a delayed effect, peaking 1-2 days after the cyclone has passed. Notably, this SST cooling exhibits a pronounced asymmetry, primarily related to the forward advection of cold wake water by geostrophic currents on the right side of the cyclone(Vincent et al., 2012).

      Taking Super Typhoon Saola (2309) as a case study, we examine the oceanic influences on typhoon-induced SST cooling through the application of three distinct models (Figures 11F-J: TSformer-w/o-aux, Figures 11K-O: TSformer, and Figures 11P-T: PSY4). Figures 11A-E illustrate the SST cooling observed by satellite microwave radiometers after the passage of Saola. Saola was classified as a tropical cyclone on August 25 in the eastern waters of Luzon, where it was notably affected by the topography of Luzon Island, which includes elevations surpassing 2000 m. The cyclone lingered in the eastern part of Luzon for four days, resulting in substantial SST cooling (see Figure 11A). On August 30, Saola crossed the Luzon Strait into the SCS, where SST

were predominantly above 27°C and exhibited a relatively uniform horizontal distribution. These conditions provided the necessary thermal energy and moisture for the further intensification of the typhoon, leading Saola to rapidly develop into a super typhoon, with the maximum SST cooling amplitude reaching approximately 4.41°C (see Figure 11B-C). The cooling effects were primarily localized on the right side of the trajectory (see Figures 11C-E). Subsequently, after September 2, Saola weakened into a tropical cyclone due to friction with the nearshore topography, coinciding with the arrival of Severe Typhoon Haikui (2311) in eastern Taiwan, which caused a decline in SST in the Taiwan Strait.

Based on the spatial distribution of SST response to Super Typhoon Saola, the performance of three models was evaluated (Figure 11). The TSformer-w/o-aux model, which lacks key drivers such as surface wind fields, inadequately simulated the typhoon-induced SST cooling, with a cooling intensity of only 1.22°C (see Figure 11I). The TSformer model, which incorporates daily auxiliary input data, accurately predicted the SST cooling characteristics, especially in the region predominantly situated to the right of the track, demonstrating a closer alignment with observational data; however, it underestimated the cooling amplitude, with a recorded intensity of 2.96°C (Figure 11M). This underestimation might be attributed to the TSformer model currently only reliance on auxiliary datasets with larger time and spatial intervals for surface forcing, a limitation discussed in section 4.3. Conversely, the PSY4 model overestimated the SST cooling intensity, with a value of 6.2°C (Figure 11R).

Notably, under the initial conditions of a local weak cooling characteristic in the Taiwan Strait (1.14°C, see Figures 11F and 11K), both the TSformer-w/o-aux and the TSformer models successfully simulated the SST cooling process induced by the new typhoon Haikui (2311) within the next five days. However, the spatial distribution of SST cooling forecasted by the TSformer-w/o-aux model was more concentrated, whereas the TSformer model provided a more accurate spatial distribution of SST cooling, closely matching satellite observations.

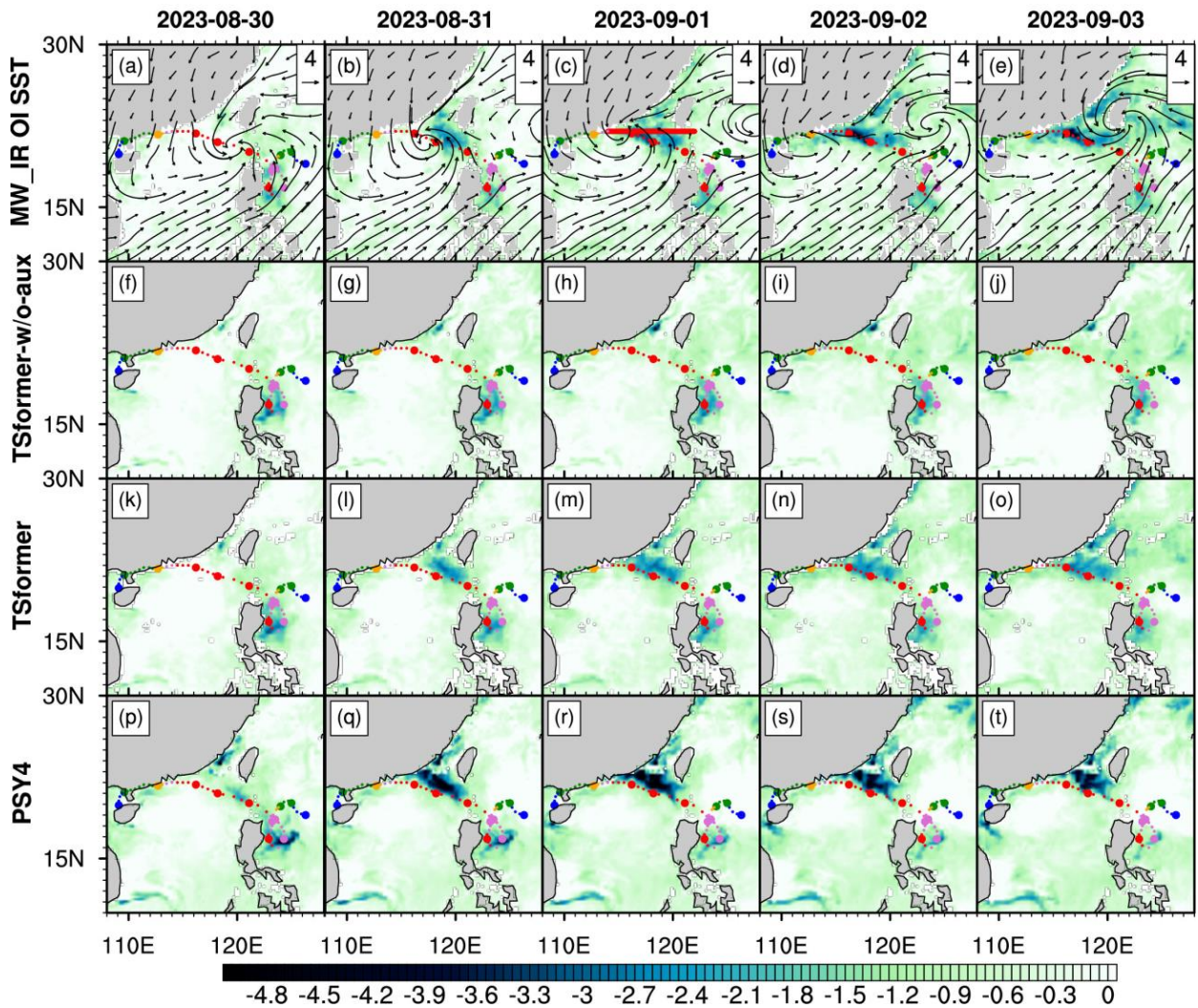

**Figure 11.** The oceanic influences on typhoon-induced cooling , as observed through MW_IR OI SST (panels A-E) and assessed using three distinct models (panels F-J: TSformer-w/o-aux, panels K-O: TSformer, and panels P-T: PSY4). In this study, SST cooling is operationally defined by comparing the average SST from August 17 to 20,2023 designated as the pre-typhoon baseline (prior to Typhoon Saola, which occurred on August 23,2023), with the SST values at each grid point from August 30 to September 4. The cooling at each grid point is quantified at the difference between these initial and subsequent average SST.

Furthermore, the influence of typhoons on the ocean is not limited to the sea surface, they can also impact the subsurface layers of the ocean through mechanisms such as near-inertial oscillations, Ekman pumping, and ocean mixing, with these effects reaching depths of approximately 60 meters (Karnauskas et al., 2021). Our study focused on the region where SST cooling was most significant, as indicated by the red line in Figure 11C, and conducted a vertical slice analysis to assess the subsurface impact of typhoons. Upon evaluating the outcomes from the three models (Figure 12), it became clear that the TSformer-w/o-aux model, due to its limited capacity to capture the characteristics of typhoon wind field changes, led to a weaker cooling and slower mixing response. Conversely, the TSformer model, which incorporates surface auxiliary observational data, effectively replicated the vertical cooling and mixing effects induced by Typhoon Saola, achieving a rapid cooling mixing depth of 80 meters within a 2-day forecast lead time (Figure 12G). The temporal and spatial distribution patterns of the vertical cooling process from the TSformer model closely matched those of the PSY4 model.

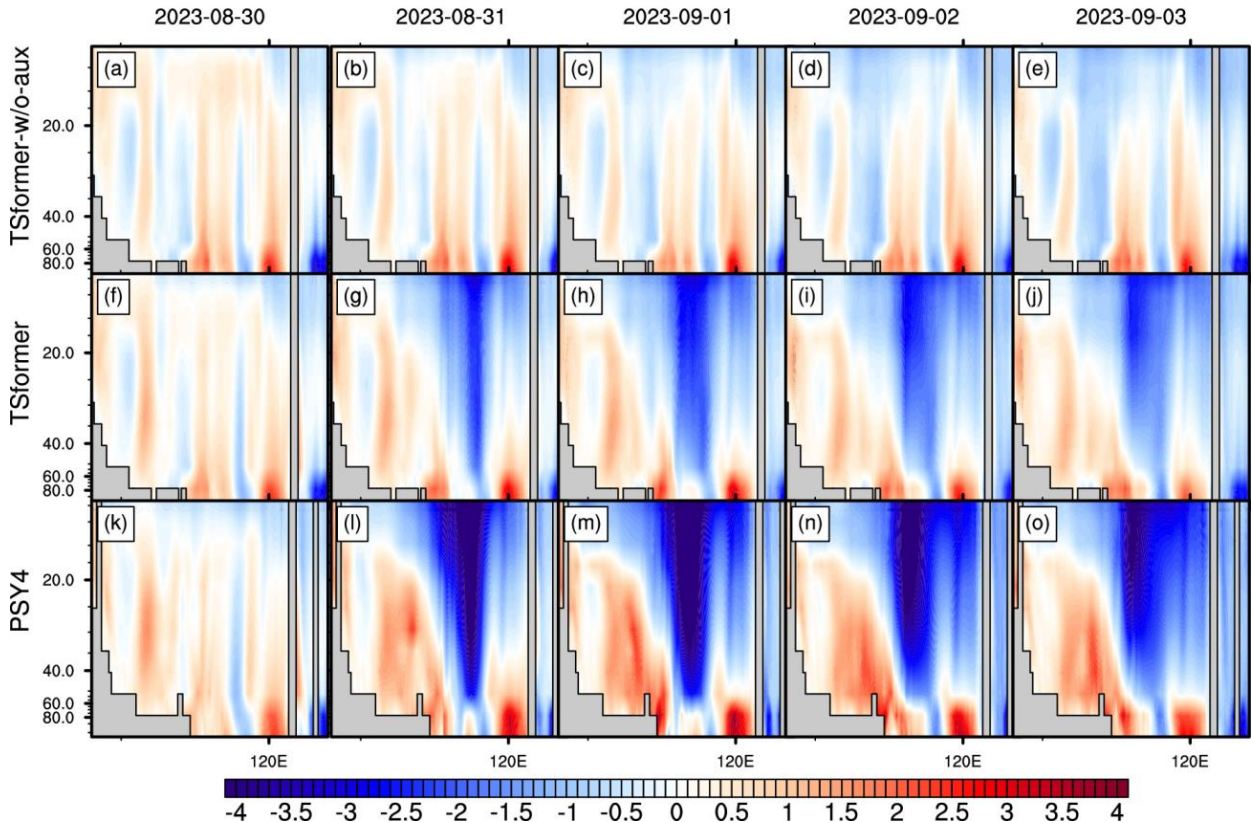

**Figure 12.** The impact of Typhoon Saola on the subsurface ocean layers at the latitude of 22N, where the SST cooling effect was most pronounced. The study presents the results from three models: TSformer-w/o-aux (panels A-E), TSformer (panels F-J), and PSY4 (panels K-O).

## 5 Conclusions

In this paper, we explore the large-scale training of an ocean forecast model utilizing the 3D ocean reanalysis product. Specifically, we adopt a hierarchical U-net encoder-decoder architecture, integrated with 3D Swin Transformer blocks, which process spatiotemporal patches of 3D TS variables and 2D surface forcing. Our target model, TSformer, is capable of forecasting 30 days of 3D eddy-resolving ocean physical variables in a non-autoregressive approach, with a daily temporal resolution and a 1/12° spatial resolution that covers 3D TS variables across 26 vertical levels.

The performance of the TSformer model has been comprehensively evaluated through its comparison with the GLORYS12V1 reanalysis data and verification against data from Argo profiling floats and satellite observations. Based on the near-real-time operational forecast results from 2023, the TSformer, which is differs from other autoregressive models, has expanded the scope of local attention computation from spatial to spatiotemporal. This expansion not only preserves the consistency of 3D TS in the physical motion process within space but also maintains long-range coherence and stability in long-term forecasts, significantly reducing cumulative errors. Moreover, the TSformer model jointly extracts both 3D TS features and 2D surface forcing characteristics through its 3D Swin Transformer modules, which are adept at handling self-attention computations in parallel at the cubic level. As a result, the TSformer has demonstrated its effectiveness in managing extreme events, exemplified by its successful forecasting of the SST cooling induced by Super Typhoon Saola. It is particularly remarkable that the TSformer has outperformed the PSY4 model in forecasting the thermocline dynamics below 20 meters depth in the SCS, a critical factor for enhancing our comprehension of the internal structure and processes of ocean. Specifically, the TSformer model is capable of completing a comprehensive 30-day forecast in approximately 40 seconds using only CPU resources, which is significantly faster by orders of magnitude compared to traditional numerical forecast models.

While the TSformer model offers computational efficiency and high accuracy in ocean eddy-resolving forecasting, there is room for enhancement. The first one is that TSformer relies on 2D daily remote sensing data for surface forcing, omitting essential parameters at the air-sea interface, such as air temperature, pressure, fluxes, and precipitation. This deficiency limits the ability to fully simulate vertical and horizontal exchanges and interactions, leading to weaker typhoon-induced cooling. To address this, future iterations of the TSformer model could integrate a broader spectrum of weather parameters with ERA5 atmospheric reanalysis datasets, and increase the capability of resolving fine-scale processes with high-resolution satellite observations (e.g., the Surface Water and Ocean Topography mission). The second one is that the deterministic nature of TSformer limits its capacity to provide probabilistic forecasts and uncertainties, especially over extended forecast periods. Enhancing TSformer to include probabilistic forecasting could mitigate these limitations by offering a spectrum of potential outcomes and their probabilities. This would enable the model to present multiple future scenarios, thereby enhancing the forecast capabilities for extreme events. By incorporating these improvements, the TSformer could achieve a more comprehensive and nuanced understanding of complex atmospheric and oceanic phenomena, ultimately refining its forecasting prowess.

## Acknowledgments

This research was financially supported by the National Key Research and Development Program of China under grant number 2021YFC3101602, the National Natural Science Foundation of China under grant number 42176017 and 41976019. We would like to extend our appreciation to the Copernicus Marine Service for providing access to the Copernicus Marine Environment Monitoring Service global ocean 1/12° physical reanalysis GLORYS12V1 dataset.Microwave OI SST data are sponsored by National Oceanographic Partnership Program (NOPP) and the NASA Earth Science Physical Oceanography Program. CCMP Version-3.1 vector wind analyses are produced by Remote Sensing Systems. Both OI SST and CCMP Data are available at www.remss.com. CMA-STI Best Track Dataset for Tropical Cyclones over the western North Pacific is obtained from www.typhoon.gov.cn. The T-S profile observation data were obtained from the ARGO global data center (ftp://ftp.ifremer.fr). The availability of this dataset was instrumental in carrying out our analysis and advancing our understanding in this field.